%% file: main.tex
\documentclass[10pt,conference]{IEEEtran}
\usepackage{cite}
\usepackage[hyphens]{url}
\usepackage{textcomp}
\usepackage{fancyhdr}
\usepackage[hidelinks]{hyperref}
\usepackage{xcolor}
\usepackage[keeplastbox]{flushend}
\usepackage[position=bottom]{subfig}
\usepackage{enumitem}
\usepackage[normalem]{ulem}

\usepackage{xcolor}
\usepackage{graphicx}
\usepackage{xparse}
\usepackage{booktabs}
\usepackage{titlesec}
\usepackage{amsmath,amssymb,amsfonts}
\usepackage{float}
\usepackage{pgfplots}
\usepackage{algorithmicx}
\usepackage{booktabs}
\usepackage{multirow}
\usepackage{xcolor}
\usepackage{graphicx}
\usepackage{tabularx}
\usepackage{adjustbox}
\usepackage{graphicx}
\usepackage{comment}
\usepackage{soul}
\pgfplotsset{compat=1.18}

\usepackage[linesnumbered,ruled,lined]{algorithm2e}
\SetKwInput{KwInput}{Input}                %
\SetKwInput{KwOutput}{Output}              %
\usepackage{dutchcal}

\usepackage{listings}
\definecolor{codegreen}{rgb}{0,0.6,0}
\definecolor{codegray}{rgb}{0.5,0.5,0.5}
\definecolor{codepurple}{rgb}{0.58,0,0.82}
\definecolor{backcolour}{rgb}{1,1,1}

\lstdefinestyle{mystyle}{
  backgroundcolor=\color{backcolour}, commentstyle=\color{codegreen},
  keywordstyle=\color{magenta},
  numberstyle=\tiny\color{codegray},
  stringstyle=\color{codepurple},
  basicstyle=\ttfamily\footnotesize,
  breakatwhitespace=false,         
  breaklines=true,                 
  captionpos=b,                    
  keepspaces=true,                 
  numbers=left,                    
  numbersep=5pt,                  
  showspaces=false,                
  showstringspaces=false,
  showtabs=false,                  
  tabsize=2,
}

\newcommand{\projname}{Pac-Sim}

\newcommand{\avgspeedupparallel}{\AvgSpeedupNaturalParallel{}}
\newcommand{\avglooppointspeedupparallel}{150.97$\times$}

\newcommand{\AVGErrorNatural}{1.63\%}

\newcommand{\AVGSPEEDUPDIFF}{3.65\%}

\newcommand{\AvgSpeedup}{26.09$\times$}
\newcommand{\MaxSpeedup}{123.32$\times$}
\newcommand{\MaxSpeedupNaturalParallel}{523.5$\times$}
\newcommand{\AvgSpeedupNaturalParallel}{210.3$\times$}
\newcommand{\AvgSpeedupNatural}{\AvgSpeedup}
\newcommand{\MaxSpeedupNatural}{\MaxSpeedup}
\newcommand{\DynamicAvgSpeedup}{6.29$\times$}
\newcommand{\DynamicMaxSpeedup}{43.96$\times$}
\newcommand{\DynamicAvgError}{3.81\%}
\newcommand{\DynamicFreqmError}{11.43\%}

\title{\textit{\projname{}}: Simulation of Multi-threaded Workloads using Intelligent, Live Sampling}

\makeatletter

\newcommand{\linebreakand}{%
  \end{@IEEEauthorhalign}
  \hfill\mbox{}\par
  \mbox{}\hfill\begin{@IEEEauthorhalign}
}

\makeatother

\author{
\IEEEauthorblockN{Changxi Liu$^{\dag}$} \IEEEauthorblockA{National University of Singapore} \and \IEEEauthorblockN{Alen Sabu$^{\dag}$} \IEEEauthorblockA{National University of Singapore} \and \IEEEauthorblockN{Akanksha Chaudhari} \IEEEauthorblockA{University of Wisconsin-Madison} \linebreakand \IEEEauthorblockN{Qingxuan Kang} \IEEEauthorblockA{National University of Singapore}\and \IEEEauthorblockN{Trevor E. Carlson} \IEEEauthorblockA{National University of Singapore}
}

\begin{document}

\maketitle

\thispagestyle{plain}
\pagestyle{plain}

\begin{abstract}
    \input{sections/0_abstract.tex}

\end{abstract}
\def\thefootnote{$\dag$}\footnotetext{Changxi Liu and Alen Sabu contributed equally to this work.}\def\thefootnote{\arabic{footnote}}
\section{Introduction}
    \label{sec:intro}
    \input{sections/1_introduction}

\section{Simulating Modern Architectures}
    \label{sec:motivation}
    \input{sections/3_motivation}
    
\section{The \projname{} Methodology}
    \label{sec:method}

\input{sections/4_methodology}

\section{Experimental Setup}
    \label{sec:setup}
    \input{sections/6_experimental_setup}

\section{Evaluation}
    \label{sec:results}
    \input{sections/7_results}

\section{Related Work}
    \label{sec:relatedwork}
    \input{sections/8_related_work}
    
\section{Conclusion}
    \label{sec:conclusion}
    \input{sections/90_conclusion}

\bibliographystyle{IEEEtranS}

\bibliography{references}

\end{document}

%% file: sections/0_abstract.tex
High-performance, multi-core processors are the key to accelerating workloads in several application domains.
To continue to scale performance at the limit of Moore's Law and Dennard scaling, software and hardware designers have turned to dynamic solutions that adapt to the needs of applications in a transparent, automatic way.
For example, modern hardware improves its performance and power efficiency by changing the hardware configuration, like the frequency and voltage of cores, according to a number of parameters such as the technology used, the workload running, etc.
With this level of dynamism, it is essential to simulate next-generation multi-core processors in a way that can both respond to system changes and accurately determine system performance metrics. Currently, no sampled simulation platform can achieve these goals of dynamic, fast, and accurate simulation of multi-threaded workloads.

In this work, we propose a solution that allows for fast, accurate simulation in the presence of both hardware and software dynamism.
To accomplish this goal, we present \projname{}, a novel sampled simulation methodology 
for fast, accurate sampled simulation that requires no upfront analysis of the workload.
With our proposed methodology, it is now possible to simulate long-running dynamically scheduled multi-threaded programs with significant simulation speedups even in the presence of dynamic hardware events. We evaluate \projname{} using the multi-threaded SPEC CPU2017, NPB, and PARSEC benchmarks with both static and dynamic thread scheduling. The experimental results show that \projname{} achieves a very low sampling error of \AVGErrorNatural{} and \DynamicAvgError{} on average for statically and dynamically scheduled benchmarks, respectively. \projname{} also demonstrates significant simulation speedups as high as \MaxSpeedupNaturalParallel{} (\AvgSpeedupNaturalParallel{} on average) for the train input set of SPEC CPU2017.

%% file: sections/1_introduction.tex
Computer architecture research heavily relies on simulations for design space exploration. 
However, microarchitectural simulation can become extremely time-consuming, particularly as the complexity of modern architectures has increased over time. 
This is especially true in the post-Dennard era, where architectures are rapidly evolving to incorporate complex dynamic optimization techniques at both the hardware and software levels to improve system performance gains at runtime. Hardware-based dynamic techniques such as dynamic cache reconfiguration~\cite{mittal2013flexiway,mittal2013master, albonesi-cache}, DVFS~\cite{eyerman2011fine,canturk2006analysis,kim2008system}, TurboBoost~\cite{charles2009evaluation} and power management~\cite{bircher2010predictive} techniques trigger optimizations based on dynamically identified hardware states (such as core frequency, cache reuse distance, etc.) to improve both energy-efficiency and overall performance of the system. Similarly, runtime information at the software level can be used to dynamically optimize code execution, to further enhance the system performance. Some of the recent efforts on software-based optimization focus on dynamically scheduling tasks among threads~\cite{duran2011ompss,diavastos2017switches} to ensure efficient resource utilization and employing just-in-time (JIT) compilation techniques~\cite{Mojo,kulkarni2019pliant,voss2001high,you2022vectorizing} that generate high-performance instructions to optimize program execution online. However, since these techniques utilize dynamic system state information in order to deploy optimizations at runtime, the execution behavior of an application (and, therefore, its performance) may vary vastly across multiple executions. 
Such variability can make it extremely difficult to determine the performance of a given workload using existing simulation methodologies. 

Conventionally, sampled simulation has served as a reliable and efficient technique to accelerate the performance estimation of multi-threaded workloads. 
In order to achieve these results, most prior works relied on either (i) profile-driven sampling~\cite{2002Sherwood, 2014carlson, sabu2022looppoint}
or (ii) statistical sampling~\cite{2003Wunderlich, PCantorSim}. 
Profile-driven sampled simulation methodologies such as SimPoint~\cite{2002Sherwood}, BarrierPoint~\cite{2014carlson}, and LoopPoint~\cite{sabu2022looppoint} split the execution of an application into a series of repeatable regions and cluster them based on their execution features. A representative element from each cluster is then analyzed or simulated in order to extrapolate the performance of the entire application. 
However, these methodologies incur a significant cost in terms of the preprocessing effort that is needed to identify representative regions. These costs include the time required to profile and cluster the execution features of all application regions, along with the storage required.  
While it has been previously argued that these costs are a one-time investment and will be amortized over multiple runs, this argument does not necessarily hold for systems that optimize code execution dynamically. 
In such cases, the program execution paths followed by an application may vary considerably due to changes of hardware and software parameters that are being optimized.
Therefore, the profiling information collected for one specific run would not necessarily extend to the program execution paths followed in the subsequent runs. 

On the other hand, sampled simulation methodologies such as SMARTS~\cite{2003Wunderlich} and PCantorSim\cite{PCantorSim} rely on statistical sampling techniques to speed up simulation-based performance measurements while meeting a given error bound. Unlike profile-driven sampling, these methodologies require minimal preprocessing and do not rely on the reproducibility of program execution paths. They are thus applicable to dynamically optimized systems. However, the simulation speedups achieved using these techniques are considerably lower than the profile-driven counterparts, and adjusting settings to achieve higher performance could lead to high errors.

For the above-mentioned reasons, it becomes challenging to sample and simulate generic multithreaded applications for dynamic hardware and software using existing methodologies. Architects need a simulation methodology that can dynamically adapt to changes in the system at runtime while accurately estimating the application's performance without relying on the reproducibility of its execution. To this end, we propose
\projname{}, a novel sampled simulation methodology that can, at runtime, efficiently analyze and sample the application to select the representative regions to be simulated in detail.
The result is a methodology that enables both fast and accurate performance evaluation without the need for up-front analysis. We accomplish this by making use of an intelligent online\footnote{We use the terms ``online'' and ``offline'' to distinguish between events that occur during and prior to the simulation of an application, respectively.} predictor and classifier that quickly and accurately decides whether the upcoming region needs to be simulated in detail. 

In short, we make the following contributions:

\begin{enumerate}[label=\roman*.,noitemsep,nolistsep,]
    \item We propose \projname{}, a methodology that goes beyond prior sampled simulation techniques to be the first to allow for dynamic hardware and software support. 
    The methodology requires no upfront analysis and relies on an online predictor for sampling decisions enabling the fast analysis of co-designed workloads. We will open-source the simulation framework for \projname{} upon acceptance.
    \item We experimentally demonstrate that \projname{} consistently improves performance in terms of speedup and accuracy over prior works that use offline profiling. We also quantify the performance benefits obtained by \projname{}, showing that they outweigh its online analysis overheads.
    \item We provide an extensive evaluation of \projname{} using standard benchmarks to compare against prior works and demonstrate best-in-class accuracy (average error of \AVGErrorNatural{}). For the SPEC CPU2017 benchmarks (train inputs) running eight threads, we show
    a maximum serial speedup of \MaxSpeedupNatural{} (\AvgSpeedupNatural{} on average) and a maximum parallel speedup of \MaxSpeedupNaturalParallel{} (\AvgSpeedupNaturalParallel{} on average).
    \item Finally, we showcase several case studies demonstrating
    that \projname{} is applicable to a number of research scenarios, including (but not limited to) the investigation of optimization techniques such as dynamically scheduled software, and improving research into dynamic hardware and hardware-software co-design.
\end{enumerate}

The rest of the paper is organized as follows. In Section~\ref{sec:motivation}, we discuss the relevant background and the challenges involved in the simulation of dynamic applications on modern architectures. Section~\ref{sec:method} presents the \projname{} methodology in detail. We then describe the experimental infrastructure in Section~\ref{sec:setup}, followed by an extensive evaluation of \projname{} in Section~\ref{sec:results} along with case studies to demonstrate the applicability of the proposed methodology. Finally, we conclude the paper in Section~\ref{sec:conclusion}.

%% file: sections/3_motivation.tex
In this section, we provide the necessary background of sampled simulation. We also discuss the challenges in simulating modern workloads and how the existing sampling methodologies are insufficient to address them.

\textbf{Sampling Single-threaded Workloads.} 
Sampling and workload reduction techniques are extensively utilized in computer architecture research for the purpose of program characterization and to reduce simulation time. Sampling methodologies allow for the evaluation of a subset of the workload (a representative sample) in detail that can be used to reconstruct the performance of the whole workload accurately. 
These methodologies split the workload into different regions (or slices) based on predetermined conditions in order to identify a representative sample. 
Prior works that explored CPU workload sampling, like SimPoint~\cite{2002Sherwood} and SMARTS~\cite{2003Wunderlich}, tend to utilize fixed instruction counts to determine regions. 
However, instruction count-based techniques could lead to inconsistent and, therefore, invalid regions~\cite{alameldeen2003hpca,2006alameldeen,2013carlson}. 
Some previous works~\cite{2006lau,yount2015graph} proposed software phase markers that identify procedure and loop boundaries that correlate with phase changes to mark region boundaries instead of using fixed-sized regions.  
On the other hand, statistical sampling methods, including SimFlex~\cite{2006wenisch}, were proposed for multi-processor throughput workloads and use instruction count for statistical sampling, but these works do not appear to be generally extensible to synchronizing multi-threaded workloads~\cite{2013carlson}. 
In the presence of synchronizing threads, the application performance tends to vary more frequently, and the statistical confidences assuming Gaussian performance distribution, as shown in SMARTS or SimFlex, may not be applicable there~\cite{chen2012statistical}.

\textbf{Sampling Multi-threaded Workloads.} Time-based sampling methodologies~\cite{2013carlson,2013ardestani} are the first to address the problem of sampling synchronizing multi-threaded applications. These methodologies, however, are slow, and as a result, they are not practical for handling realistic workloads. 
On the other hand, methodologies like BarrierPoint~\cite{2014carlson}, TaskPoint~\cite{2016grass}, and LoopPoint~\cite{sabu2022looppoint} select specific program constructs, such as barrier synchronization primitives, task instances, and loops, respectively, to identify periodic behavior. This enables the utilization of representative-sized regions for simulation, regardless of the program's length. 

\textbf{Feature Vectors.} Profiling captures feature vectors up-front to characterize the execution behavior of an application across regions. Previous works have introduced several microarchitecture-independent feature vectors, of which basic block vectors (BBVs)~\cite{2002Sherwood, 2006perelman} are the most widely used for performance characterization. Lau et al.~\cite{lau2005} showed a strong correlation between BBVs and region performance. Apart from BBVs, Shen et al.~\cite{shen2004locality} introduced LRU stack distance vectors (LDVs)~\cite{mattson1970ldv} to summarize program behavior for different regions. BarrierPoint~\cite{2014carlson} combines BBVs and LDVs into a signature vector (SV) in an attempt to represent more accurate features of multi-threaded applications. 
Furthermore, Cotson~\cite{argollo2009cotson} and Dynamic sampling~\cite{falcon2007combining} record statistics such as the number of instructions executed, memory accesses, exceptions, bytes read or written, etc., in order to plot the feature of a given region. 
Unfortunately, none of these offline techniques are capable of representing runtime optimizations adopted by applications.

\begin{figure}[t]
    \centering
    \includegraphics[width=\linewidth]{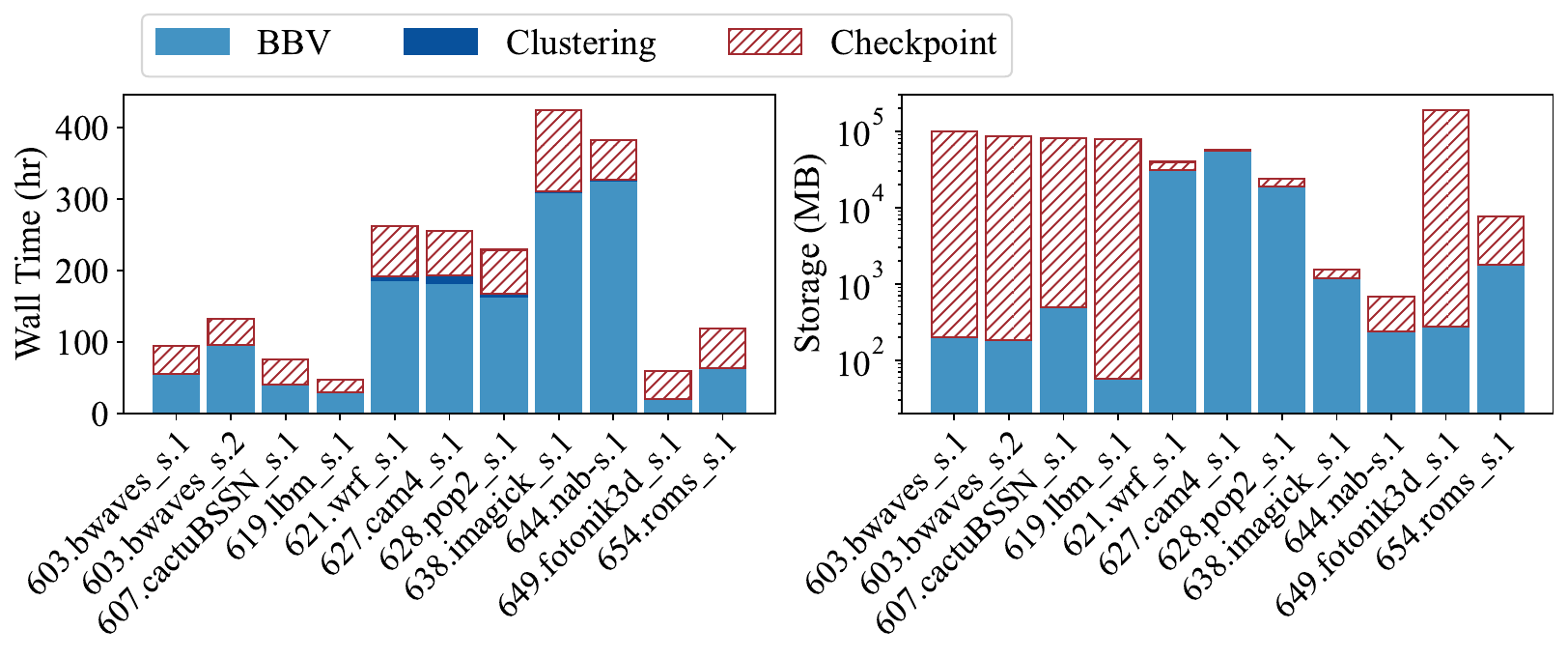}
    \caption{The figure shows the resource utilization of a recent multi-threaded sampled simulation technique, LoopPoint, 
    for the SPEC CPU2017 benchmarks with the \texttt{ref} inputs running eight OpenMP threads. The graph on the left shows the time required to generate the profiling data (with checkpoints stored as pinballs~\cite{2014harish}), whereas the graph on the right shows the amount of storage required. 
    }
    \label{fig:resource}
\end{figure}

\textbf{Overheads.} 
Figure~\ref{fig:resource} illustrates the overhead of profiling data for LoopPoint (the evaluation was performed using the LoopPoint tools~\cite{lpgithubweb}) methodology, indicating that profile-driven methodologies incur significant overheads. When it is required to emulate an architecture (for example, simulating ARM or RISC-V binaries on x86) during profiling, it is necessary to resort to functional simulation to gather feature vectors, which can be a time-consuming process. For instance, Sandberg et al.~\cite{sandberg2015fsa} demonstrated that it took up to a month to generate profile data for SPEC CPU2006 benchmarks using simulators like gem5.
In addition, profiling for asymmetric hardware, such as the \textit{big.LITTLE} cores is challenging as the operating frequency (and other dynamic hardware settings) of each core is unknown at the time of profiling.
Handling and storing simulation checkpoints can be a daunting task. For example, x86 checkpoints like ELFies~\cite{patil2021elfies} require a significant amount of storage space. These checkpoints are usually specific to the simulator being used, and they are often tied to a particular software version or hardware configuration.

\textbf{Hardware and Software Dynamism.}
Researchers have introduced several dynamic optimization techniques in hardware and software to achieve higher performance and reduce power consumption. Techniques such as dynamic voltage and frequency scaling (DVFS) and cache reconfiguration have been developed to adjust the hardware state in response to executed instructions and active processes. 
Software optimization techniques~\cite{Mojo,kulkarni2019pliant,voss2001high,you2022vectorizing} generate high-performance instructions at runtime. 
Additionally, dynamic scheduling techniques~\cite{duran2011ompss} have been developed for multi-threaded applications. In such cases, profile-driven sampling methodologies result in different performances for each execution. Methodologies such as trace-based simulations~\cite{butko2015trace} or deterministic replay platforms~\cite{2010harish} can guarantee consistent performance across multiple executions but demand extensive profiling and large storage resources. 
Dynamic hardware events, such as changes in core frequency, cache size, etc., are unknown during profiling. These events are performance or power-dependent and are usually hard to predict.
Sherwood et al.~\cite{sherwood2003phase} utilize a Markov Predictor to predict the phase behavior at runtime. Kihm et al.~\cite{kihm2007phase} propose switching to the detailed simulation mode whenever the BBV variance exceeds a specified threshold. But these methods work only for single-threaded applications as the phase behavior of synchronizing multi-threaded applications varies frequently due to the interaction of threads.

\textbf{Requirements for Fast and Accurate Simulation. }
Sampled simulation without upfront analysis is promising under these dynamic software and hardware constraints.
Therefore, it is imperative to leverage the best aspects of SimPoint-like and SMARTS-like methodologies to achieve optimal simulation efficiency and accuracy.
In this work, we incorporate application analysis to guide sampled simulations, similar to SimPoint-like methodologies but without 
the need for upfront pre-processing, as seen in SMARTS-like methodologies. 
Instead, we make intelligent simulation decisions through online learning. Moreover, the proposed methodology can accommodate hardware state changes, software features, and other factors that affect simulation results. It is essential to implement efficient and lightweight online profiling, clustering, and warmup techniques for optimal performance.
Therefore, to quickly estimate the performance of multithreaded applications running on next-generation dynamic hardware and software, a sampled simulation methodology is needed that can dynamically adapt to changes in the system at runtime while accurately determining relevant performance metrics.

%% file: sections/4_methodology.tex
\begin{figure}[t]
    \centering
    \includegraphics[width=\linewidth]{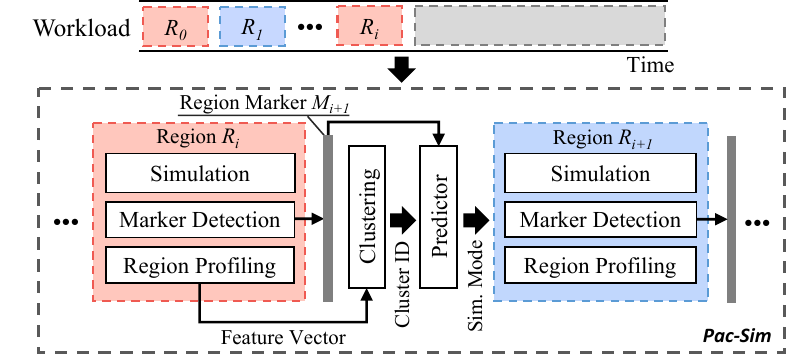}
    \caption{Figure depicts the workflow of \projname{}. Consider a multi-threaded workload with regions till $R_{i}$ are identified (as shown above). 
    First, \projname{} monitors the application code structure to determine an appropriate region marker $M_{i+1}$, which marks both the end of the region $R_{i}$ and the start of the region $R_{i+1}$. 
    Next, the feature vector and simulation results for $R_{i}$ are collected, and a prediction mechanism 
    determines the simulation mode for region $R_{i+1}$. Finally, region $R_{i+1}$ will be simulated, either in detail or in fast-forward mode.}
    \label{fig:overview}
\end{figure}

In this section, we describe our proposal for an end-to-end sampled simulation methodology, \projname{} (depicted in Figure~\ref{fig:overview}), that supports both dynamic hardware and software without requiring up-front workload analysis. \projname{} consists of five main stages: Marker Detection, Region Profiling, Clustering, Prediction, and Simulation, which are all carried out online. We have carefully designed each of these stages to minimize the runtime overhead of the methodology while maintaining the sampling accuracy. 
An important advantage of an online sampled simulation methodology like \projname{} is its ability to accurately determine the execution profile of an application without relying on the reproducibility of a program's execution paths. This characteristic allows \projname{} to accurately analyze and evaluate dynamic multi-threaded applications, accounting for any performance variability that may occur at runtime.

\projname{} operates by making use of the program structure and runtime hardware state to identify the regions and their boundaries online. Each of these region boundaries or \textit{markers} defines the ending of the current region and the beginning of the next region (Section~\ref{sec:sampled}). Once a marker is identified, \projname{} collects the profiling data and simulation results of the current region (Section~\ref{sec:profiling}) and clusters it with the previously identified regions to determine its cluster ID (Section~\ref{sec:clustering}). This cluster ID is added to the program execution history, which is then used by the \textit{Predictor} (Section~\ref{sec:prediction}) along with the current marker and hardware state to predict whether the next region needs to be simulated in detailed mode or fast-forward mode.

While we only demonstrate the effectiveness of \projname{} in estimating the performance of synchronizing multi-threaded workloads in this work, our methodology has the potential to support a variety of modern workload classes, such as cloud and mobile applications, and could also be implemented for full system simulations. However, in such cases, various factors must be taken into consideration, such as kernel and driver performance, which can significantly impact the overall efficiency of the workloads. In this work, we focus on user-space workloads, and enabling support for the above-mentioned use cases is out-of-scope in this context which we leave for future work.

\subsection{Online Region Detection}
\label{sec:sampled}

Previous research~\cite{2006lau,2014carlson,sabu2022looppoint} has shown that certain program constructs, such as barriers or loops, can be utilized to characterize the phase behavior of multi-threaded applications by splitting them into a series of individually analyzable regions. Since barriers represent the global synchronization points within a program execution, all threads align at these points, making them natural boundaries for application regions. However, relying solely on barriers to split an application may not be ideal, especially in the presence of large inter-barrier regions, as this can lead to low simulation speedups as representatives can still be too large to complete detailed simulation in a reasonable amount of time. In contrast, loops offer a finer level of granularity, allowing for greater control over the size of regions.
Typically, multi-threaded applications consist of both loops and barriers in varying proportions. 
The online \textit{Marker Detector} combines both of these program constructs to effectively split multi-threaded applications into regions with sizes that are well-suited for clustering while also avoiding aliasing~\cite{casas2012frequency}.
The Marker Detector uses the following approach in order to identify the barrier- and loop-based markers online:

\textbf{Barriers.} Typically, a multi-threaded region begins with a \texttt{fork} call, which spawns additional worker threads and ends with a \texttt{join} call, which terminates the current thread and synchronizes with other threads. A new region is triggered at events of thread creation and termination, as regions with different active threads have different performances. For multi-threaded programs that use the OpenMP library, special function names are generated depending on the compiler used. We utilize this information in the online Marker Detector to quickly and efficiently detect barriers without much overhead. 

\textbf{Loops.} Both loop and conditional statements use conditional branch instructions, with the target address usually given as an offset from the instruction pointer. The key difference between the two statements is that the offset of the branch instructions in a loop statement is usually negative, whereas that in a conditional statement is positive. While there are exceptions, it is generally sufficient to select conditional branches with negative offsets as markers for loops. We also make sure to disregard spinloops from our analysis. 

As an application executes, the Marker Detector identifies markers online, splitting the application into multiple regions. While doing so, it also monitors the region sizes to ensure they fall approximately within the bounds of $\delta_{min}$ and $\delta_{max}$ instructions. A minimum number of instructions, $\delta_{min}$, is necessary to capture the frequent variations in the multi-threaded program behavior and accurately cluster the obtained regions. Whenever the Marker Detector chooses barrier-based markers as region boundaries, the size of the region can be as small as $\delta_{min}$ instructions but no larger than $\delta_{max}$ instructions. 
Otherwise, the Marker Detector chooses the first loop-based marker it encounters beyond $\delta_{max}$ instructions as the next region boundary.
For loop-bounded regions, it is necessary to keep region sizes large enough to avoid aliasing~\cite{2013carlson}.
In our experiments with fixed region sizes of 10 million, 20 million, 50 million, and 100 million instructions, the SPEC CPU2017 benchmarks showed average error rates of 6.9\%, 3.3\%, 1.8\%, and 1.8\%, respectively. We set the lower bound $\delta_{min}$ to be 20 million to ensure sampling accuracy and the upper bound $\delta_{max}$ to be 50 million for better performance.

\textbf{Hardware State.} The Marker Detector also monitors the hardware state of the simulated system. If it detects changes, the current region is ended at the next marker so that each region has a consistent hardware state. Once a marker is detected, the program counter (PC) and the hardware state of the simulated system are collected and stored corresponding to the marker. The collected hardware state includes the system parameters, like processor frequency, cache size/configuration, power management techniques, etc., that can be configured during runtime.

\subsection{Online Region Profiling}
\label{sec:profiling}

Conventionally, BBVs have been used to characterize the execution behavior of code regions, as they have been shown to exhibit a strong correlation with the region's performance~\cite{lau2005}. 
BBVs record the execution counts of each basic block (i.e., code blocks with single entry and exit points) within a given code region. The number of dimensions for a BBV depends on the number of basic blocks executed, which could range anywhere from thousands to even millions for very large applications. This presents a major challenge for online analysis of BBVs as the time and effort required for this stage would significantly increase as the vector dimensionality increases. SimPoint~\cite{2002Sherwood} uses random linear projection~\cite{dasgupta2000projection} to overcome this problem. However, this method is not suitable for our online algorithm as the matrix-vector multiplication operations involved could introduce significant runtime overheads.

\begin{figure}[!t]
    \centering
    \includegraphics[width=\linewidth]{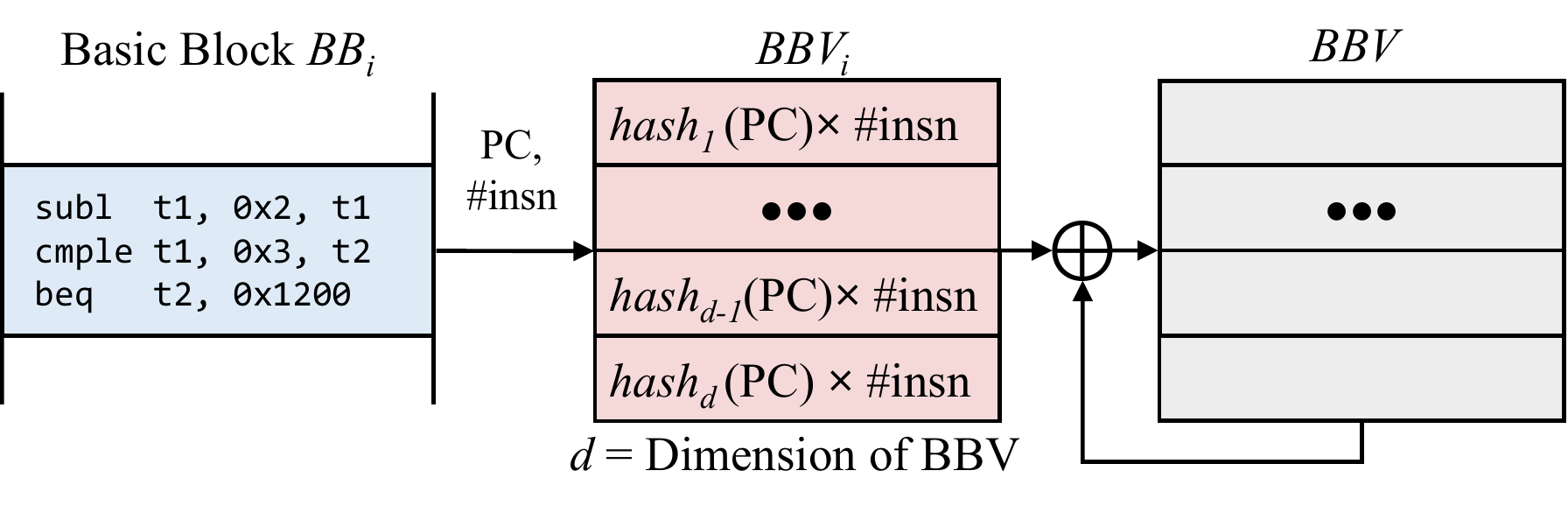}
    \caption{The figure shows the workflow of online BBV generation. Whenever a basic block $BB_i$ is encountered, a corresponding execution fingerprint $BBV_i$ is generated using hash functions applied to the program counter of $BB_i$ and the number of instructions it contains. $hash_{1}$ to $hash_{d}$ are $d$ distinct hash functions, where $d$ is the dimension of the BBV. The BBV for each region is obtained by accumulating all $BBV_i$s that belong to the region.}
    \label{fig:onlinebbv}
\end{figure}

To overcome these issues, we propose a fast online BBV generation technique (illustrated in Figure~\ref{fig:onlinebbv}). Rather than creating a fixed-size BBV for each region, we use an online projection technique to generate fixed-size vectors $BBV_{i}^{\prime}$ for each basic block $BB_{i}$, where the elements of $BBV_{i}^{\prime}$ are computed by multiplying the instruction count of a basic block with the hash results of its program counter (PC) value. We use the hash function \texttt{drand48()}, which generates pseudo-random numbers for an integer value input. The initial four dimensions of the online BBV are determined using the hash values utilizing inputs PC, PC+1, PC+2, and PC+3, respectively. The values of the subsequent four dimensions are generated using the output of the preceding four dimensions as inputs to the hash function.
We experimentally determined that using 16 dimensions adequately captures the representation of a region using the online BBV. The resultant $BBV_{i}^{\prime}$ vectors are then accumulated to obtain the per-thread BBV ($BBV_{online}^{\prime}$) for the given region, which can be represented as: 
\begin{equation*}
   BBV_{online}^{\prime} = \sum_{i}{BBV_{i}^{\prime}} = \sum_{i}{(BBV_i \cdot M_{proj})},
\end{equation*}
where the values of the elements in $M_{proj}$ are generated using hash functions as mentioned above.
This $BBV_{online}^{\prime}$ for a region is analogous to the BBV utilized in SimPoint, which is obtained through random linear projection. The projected down BBV used in SimPoint, $BBV_{offline}^{\prime}$, is obtained from the dot product of the actual BBV of the region and projection matrix $M_{proj}$: 
\begin{equation*}
 BBV_{offline}^{\prime} = BBV \cdot M_{proj} = \sum_{i}{(BBV_i\cdot M_{proj})}.
\end{equation*}

We then normalize these per-thread BBVs and concatenate them into a single global-BBV vector to represent the software feature of a given multi-threaded code region. This eliminates the need to perform computationally intensive dimensionality reduction techniques online and allows \projname{} to quickly determine the BBVs without much overhead.

\subsection{Determining Region Similarity} 
\label{sec:clustering}

\projname{} employs an online clustering mechanism to group regions with similar execution behavior based on the feature vectors collected for each region in the online profiling stage. The clustering, which is done at the end of simulating each region, is required for the learning process of the Predictor.
Prior works, like SimPoint~\cite{2002Sherwood}, cluster feature vectors using the k-means algorithm~\cite{forgy1965}. However, k-means uses an iterative refinement technique that is computationally intensive, and therefore, it would not be practical to use this algorithm for determining region similarity online. 

In order to reduce this computational load and enable real-time clustering, we devise an alternative technique for clustering feature vectors (i.e., global-BBVs) in \projname{}.
In our technique, we maintain two separate queues: (i) detailed queue and (ii) fast-forward queue. The detailed queue includes the BBVs corresponding to the regions that have been simulated in detail, while the fast-forward queue includes those corresponding to the regions that have been fast-forwarded.
When a new BBV is recorded, it is first compared with the BBVs in the detailed queue. If its distance from any of these BBVs is less than the specified threshold $\theta$, then we return the cluster ID of the closest region. If there is no region whose distance is less than $\theta$, we repeat the same procedure with the regions in fast-forward queue. If we still don't find similar regions, we assign a new cluster ID for the current region and insert it into the BBV queue corresponding to its simulation mode. 
In our experiments, we set $\theta=0.05$ to ensure a reasonable simulation accuracy while maintaining high speedups. 
To further improve the efficiency of our clustering technique, we incorporate the triangle inequality optimization~\cite{elkan2003using} into our algorithm, which can skip redundant BBV distance calculations. We consider Euclidean distance for all BBV distance calculations.

\subsection{Prediction Mechanism}
\label{sec:prediction}

\begin{figure}[t]
    \centering
    \subfloat[\label{fig:trietree}]{
    \includegraphics[width=0.74\linewidth]{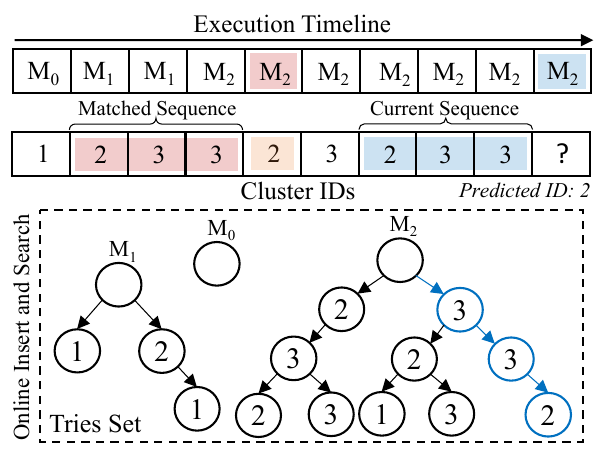}}
    \subfloat[\label{fig:predictor}]{
    \includegraphics[width=0.24\linewidth]{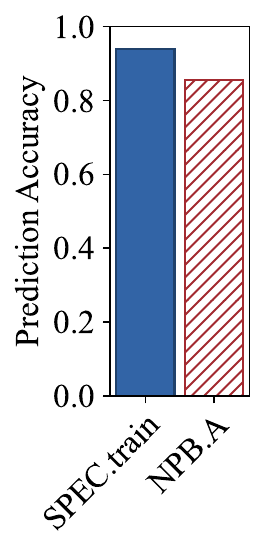}}
    \caption{The predictor utilizes the {trie}~\cite{bodon2003trie} data structure to quickly predict the cluster ID of the next region by searching for a similar history with the same region start marker $M_i$. In this example, the
    cluster ID of the next region is predicted to be $2$ since the prior region with the cluster ID of $2$ has the same start marker $M_2$ and the longest matching sequence ($2\rightarrow3\rightarrow3$). Plot (b) shows the accuracy of the predictor for different benchmark suites.}
    \end{figure}

\projname{} employs a \textit{Predictor} -- an online prediction mechanism that leverages region markers, execution history, and hardware state to predict the phase behavior of the next region in an application and decide its simulation mode at runtime. 

\textbf{Region Markers.} 
The Marker Detector identifies PC-based region markers that act as the boundaries of the regions. In certain cases, using region markers to classify regions is effective for applications where the same part of the code displays similar phase behavior, as in the case of \texttt{619.lbm\_s.1} and \texttt{644.nab\_s.1} using train inputs. 
       
\textbf{Execution History.} When executing the same part of the source code, differences in memory access patterns, branching, etc., can result in varying phase behavior at runtime. We, therefore, make use of execution history, which is a sequence of the cluster IDs of prior regions, to predict these differences in the phase behavior among applications. 
    
\textbf{Hardware State.} \projname{} takes into account the state of the simulated system, such as core frequency, while executing each region. \textit{Predictor} predicts the next region to be detailed mode if there are no prior similar regions with the same hardware state. 
The Predictor decides the cluster ID of the next region by choosing the cluster ID of the previous region  
with the same region marker and has the longest matching sequence. For the regions that do not have a previous region with the same start marker or the same history, \projname{} enables detailed execution for that region. 
Then it decides the simulation mode of the next region by checking whether prior regions with cluster ID and the current hardware state are simulated in detailed mode. This history is learned online and is updated every time \projname{} finishes simulating a region.

To accelerate this stage for large application lengths, we further optimize our clustering algorithm to reduce its average search time complexity from $\mathcal{O}(n^{2})$ to $\mathcal{O}(n)$. This is achieved by maintaining the execution history in a \textit{trie}~\cite{bodon2003trie} data structure, with a maximum depth of 16, which allows for more efficient \textit{search} and \textit{insert} operations.
In \projname{}, we utilize the trie data structure to maintain the execution history of the application being simulated and quickly predict the cluster ID of the next region based on this information. 

\begin{figure}[t]
    \centering
    \includegraphics[width=\linewidth]{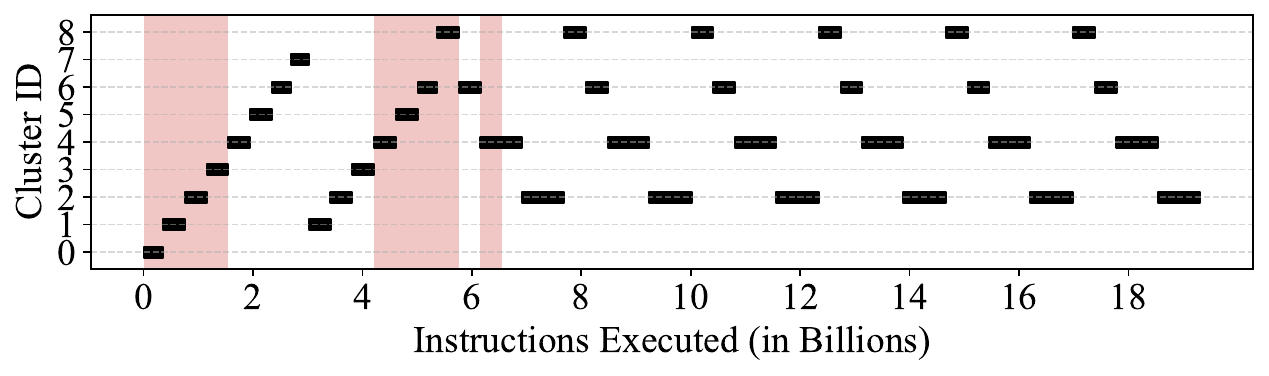}
    \caption{The graph shows the regions identified using \projname{} for the NPB benchmark \texttt{ft}, grouped together with the respective cluster they belong to. The shaded portion represents the regions that are simulated in detail.}
    \label{fig:accuracy}
\end{figure}

Figure~\ref{fig:trietree} illustrates the usage of tries to predict the cluster ID of the next region by considering the example of a hypothetical execution sequence. 
\textit{Insert:} The cluster ID of the current region is inserted into the trie. In Figure~\ref{fig:trietree}, when the online clustering of the fifth region is finished, we insert the current cluster ID $2$ for both branches corresponding to the three histories: $3$, $3\rightarrow3$, and $2\rightarrow3\rightarrow3$.
\textit{Search:} Once Insert of the current region is completed, the cluster ID of the next region is predicted by searching the trie for a matching cluster ID sequence. The search operation ends when the sequence matches one of the leaf node paths. 
Note that two regions having the same marker do not necessarily mean that the regions belong to the same cluster.

Figure~\ref{fig:predictor} shows the average accuracies of the online predictor for the benchmarks of SPEC CPU2017 and NPB are 94\% and 85\%, respectively, ensuring the sampling accuracy and performance of \projname{}. The accuracy of the predictor is determined by comparing the predicted cluster ID prior to simulating the region with the actual cluster ID obtained through clustering after simulation.
Figure~\ref{fig:accuracy} shows the results of the Predictor in clustering different regions identified by \projname{} simulating the \texttt{ft} benchmark from the NPB benchmark suite using eight threads. We observe that the majority of regions from each cluster are simulated in detail (shaded portions). This is in accordance with the learning phase of our algorithm where \projname{} works to establish a comprehensive understanding of the phase behavior of the application.

\subsection{Simulation by Application Reconstruction}
\label{sec:reconstruction}

Previously proposed multi-threaded sampling methodologies \cite{2014carlson,sabu2022looppoint} rely fully on offline analysis to determine the regions that need to be simulated in detail.
\projname{} assumes no prior knowledge about the nature of the workload that it is about to simulate. Instead, it (a) samples regions online during the simulation and (b) uses the detailed simulation results of previous regions to  
estimate the performance of the current fast-forwarded region by applying the four different methods described below successively until convergence is reached.
\begin{enumerate}[label=\roman*.,noitemsep,nolistsep,leftmargin=*]
    \item Use the detailed performance metrics of a region that belongs to the same cluster and has the same start marker as the current region.
    \item Use the detailed performance metrics of a region that belongs to the same cluster.
    \item Use the performance of a region with the closest BBV ($\theta = 1$) and the same number of active threads.
     \item Use the average performance of the regions that have the same number of active threads. 
\end{enumerate}

We extrapolate the runtime $T_{r}$ of a fast-forwarded region $r$ using the region $r^{\prime}$ by $T_{r}=T_{r^{\prime}} \frac{insn_{r}}{insn_{r^{\prime}}}$, where $T_{r^{\prime}}$ is the runtime of the previous region $r^{\prime}$ identified above, and $insn_{r}$ and $insn_{r^{\prime}}$ are the maximum instruction counts among all threads for the regions $r$ and $r^{\prime}$, respectively.

\textbf{Runtime Hardware Events.} \projname{} takes into account the state of the simulated system while estimating the performance of the fast-forwarded region. As runtime hardware events can happen at any time, we do not guarantee the regions that are divided by those events to be large enough. In such cases, we estimate the performance of these regions using the closest previous region with the same hardware state, as these regions are too small to be clustered. Moreover, the impact of these regions on the overall application performance is typically negligible as the regions are too short. 

\subsection{Sampled Simulation in Parallel}

\begin{figure}[t]
    \centering
     \includegraphics[width=\linewidth]{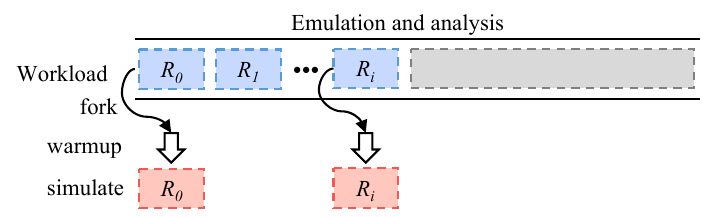}
    \caption{The workflow of \projname{} when the representative regions are simulated in parallel. \projname{} starts in the emulation mode, collecting feature vectors and MTR~\cite{barr2005accelerating} warmup data online, and then predicts the simulation mode of the next region. For regions predicted for detailed mode, \projname{} forks new processes to perform warmup and detailed simulation.}
    \label{fig:overview:parallel}
\end{figure}

Pac-Sim is primarily targeted for runtime varying scenarios using live sampling. However, for statically scheduled multi-threaded applications, Pac-Sim can support sampled simulation in parallel, similar to checkpoint-based mechanisms, to further speed up the sampled simulation. The workflow of \projname{} for parallel simulation is shown in \autoref{fig:overview:parallel}. Previous methods, like LiveSim~\cite{hassani2016livesim} and LoopPoint~\cite{sabu2022looppoint}, require offline analysis and store checkpoints for sampled simulation. A huge amount of storage is required for these methods, as mentioned in Section~\ref{sec:motivation}. Pac-Sim starts in emulation mode, collecting feature vectors and warmup data online, and then predicts the simulation mode of the next region. For regions predicted for detailed mode, Pac-Sim forks new processes, which run in parallel, to perform warmup and detailed simulation. 
\projname{} reconstructs the performance of the entire application once the whole application is emulated and the simulation of all regions is completed.

\subsection{Microarchitectural Warmup}
\label{sec:warmup}
One of the major challenges of sampled simulation is to build up the accurate microarchitectural state prior to the detailed simulation of each region. Choosing the right warmup technique that can directly build this state is crucial in order to achieve the highest speedup. Methodologies like SMARTS~\cite{2003Wunderlich} and time-based sampling techniques~\cite{2013carlson,2013ardestani} keep functional warming enabled for the entire sampled simulation leading to large slowdowns.
We find that the statistical warmup techniques~\cite{haskins2003memory,barr2005accelerating,van2006efficient,nikoleris2016coolsim} can reconstruct the accurate microarchitectural state of a simulated system online. We select MTR~\cite{barr2005accelerating} to be used with \projname{} as it can rapidly collect memory reference patterns during fast-forward mode and reconstruct the cache state before switching to detailed mode. 
Caches are larger structures compared to branch predictors or prefetchers, and hence we limit the simulation infrastructure to explicit cache warming, as the smaller structures, like prefetchers, are warmed quickly. Moreover, we maintain larger regions (minimum region sizes of 20 million instructions) to achieve good warm-up performance for other microarchitectural structures. It is also possible to increase the amount of warmup needed for different structures, and there are different ways to solve this problem, but warmup is a challenge that is workload-specific. The exploration of additional warmup scenarios is outside the scope of this work.

%% file: sections/6_experimental_setup.tex
In this section, we describe the experimental setup used to evaluate \projname{}. We begin by providing the details of the simulation framework used in our experiments. We then describe the different workloads that are used to evaluate the performance of our methodology. 

\subsection{Simulation Tools}

In this work, we use a modified version of the Sniper multi-core simulator \cite{2011carlson} (version 7.4), which is updated to support loop-based and barrier-based region specifications in order to evaluate \projname{}. Sniper is a many-core simulator using high-level abstract models and is widely used for architectural evaluation and design space exploration. Note that our methodology does not utilize any features specific to the Sniper simulator. Therefore, porting the methodology to other simulators, such as gem5~\cite{binkert2011gem5} or ZSim~\cite{2013sanchez}, should be relatively straightforward. To demonstrate that \projname{} is indeed a microarchitecture-independent methodology, we experimentally evaluate it by running simulations upon two different %
processor configurations that mimic the performance/behavior of Intel's Gainestown and Skylake\footnote{Note that Gainestown is the latest microarchitecture available on Sniper simulator that has been validated against hardware. We made modifications to the back-end of Sniper to support the contention model and instruction latencies for Skylake architecture.}~\cite{doweck2017inside} microarchitectures using Sniper. The configuration details for each of these models are listed in Table~\ref{tab:arch}.

\begin{table}[h]
\caption{The configuration parameters we used for Gainestown and Skylake microarchitectures on Sniper.}
    \centering
    \resizebox{\linewidth}{!}{%
    
    \begin{tabular}{lll}
    \toprule
    Component     &  Gainestown Parameters & Skylake Parameters \\
    \midrule
    Processor     & 1, 8 cores & 1, 8 cores    \\
    Core  &    2.66 GHz, 128-entry ROB  &   2.66 / 3.7 GHz, 224-entry ROB\\
    L1-I / L1-D       & 32KB, 4 / 8 way, LRU    &  32KB, 8 / 8 way, LRU    \\
    L2 cache   & 256KB, 8 way, LRU  &   1MB, 16 way, LRU \\
    L3 cache   & 8MB (shared), 16 way, LRU & 22MB (shared), 12 way, LRU \\    
    \bottomrule
     \end{tabular}}%
    \label{tab:arch}
\end{table}

In order to speed up the simulation, \projname{} intelligently switches among the three simulation modes supported by Sniper, namely, fast-forward mode, cache-only mode, and detailed simulation mode. The fast-forward mode is used to reach a particular point in an application during simulation without enabling the performance models. The cache-only mode performs the functional warming of the caches, whereas the detailed simulation mode is the default simulation mode that enables the timing model for performance estimation. For \projname{}, we capitalize on this split execution and timing model architecture to fast-forward in the front-end of the simulator so that the simulation wall time is further minimized.

Every time \projname{} switches from fast-forward to detailed simulation mode; the cache state is reconstructed at the beginning of the region using the memory time-stamp record~(MTR)~\cite{barr2005accelerating} technique. We implement MTR in Sniper to collect the cache line information accessed by each \textit{Load} and \textit{Store} instruction during simulation, ordered in LRU fashion per set, and then inject the requests into the cache in the correct order to rebuild the appropriate cache state.

\subsection{Benchmarks Used}
To demonstrate the wide applicability of \projname{}, we experimentally evaluate the methodology using multiple benchmark suites %
such as (i) the SPEC CPU2017 benchmark suite~\cite{2018Bucek}, (ii) the NAS Parallel Benchmarks (NPB)~\cite{bailey1995parallel} version 3.4.2, and (iii) the PARSEC~\cite{bienia2011parsec} version 3.0 benchmark suite. Note that these are multi-threaded benchmarks that synchronize frequently and share memory.

We configure these benchmarks to use two different multi-threaded programming models, namely OpenMP~\cite{OpenMP} and OmpSs~\cite{duran2011ompss}. 
OpenMP~\cite{OpenMP} provides a set of compiler directives, library routines, and environment variables that help developers to parallelize their code. On the other hand, OmpSs~\cite{duran2011ompss} extends OpenMP, and it is able to dynamically manage and schedule tasks to maximize multi-threaded application performance.
We set up the multi-threaded benchmarks to use \textit{passive} thread wait policy, meaning that the threads will sleep while waiting for other threads at a synchronization point.

SPEC CPU2017 is a collection of benchmarks used for performance evaluation in computer architecture research. 
In our experiments, we use the \textit{speed} version of multi-threaded 
SPEC CPU2017 benchmarks that are parallelized with OpenMP. 
The benchmarks are compiled using GCC 6.4.0 and GFortran with the \texttt{-O3} compiler flag for x86-64 architecture. We configure these benchmarks to run with eight threads and evaluate them using the \textit{train} input set.
NAS Parallel Benchmarks (NPB)~\cite{bailey1995parallel} is another set of benchmarks widely used to evaluate the performance of highly parallel systems in computer architecture.  The reference implementations of these benchmarks are available in the two most commonly used programming models, i.e., MPI and OpenMP. In our experiments, we use the OpenMP-based implementation with input \textit{class A} and generate the binaries using icc compiler (with \texttt{-O2} flag) as part of the Intel oneAPI (version 2022.0.2) toolkit. 
We also present experimental evaluations of \projname{} using PARSEC, which is another standard benchmark suite consisting of computationally intensive applications designed to facilitate the study of multi-core systems with shared memory. PARSEC implementations are available in both OpenMP and OmpSs~\cite{chasapis2015parsecss} versions. In our experiments, we use both these versions with the \textit{simlarge} input set.

\begin{figure}[t]
   \centering
    \includegraphics[width=\linewidth]{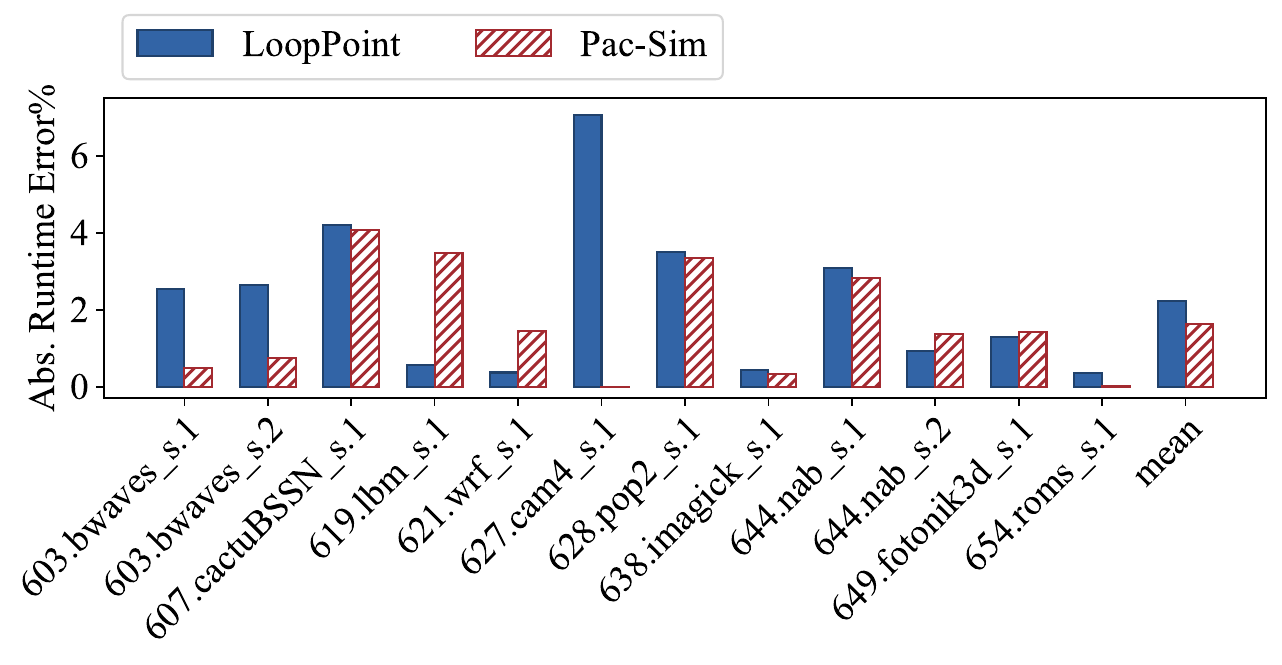}
    \caption{A comparison of the absolute runtime prediction error for LoopPoint and \projname{} for 8-threaded SPEC CPU2017 benchmarks using train inputs. \projname{} achieves similar levels of accuracy compared to LoopPoint.} 
    \label{fig:sota:err}
\end{figure}

\begin{figure*}[!t]
    \subfloat[Parallel Speedup\label{fig:sota:parspeedup}]{
    \includegraphics[width=0.48\linewidth]{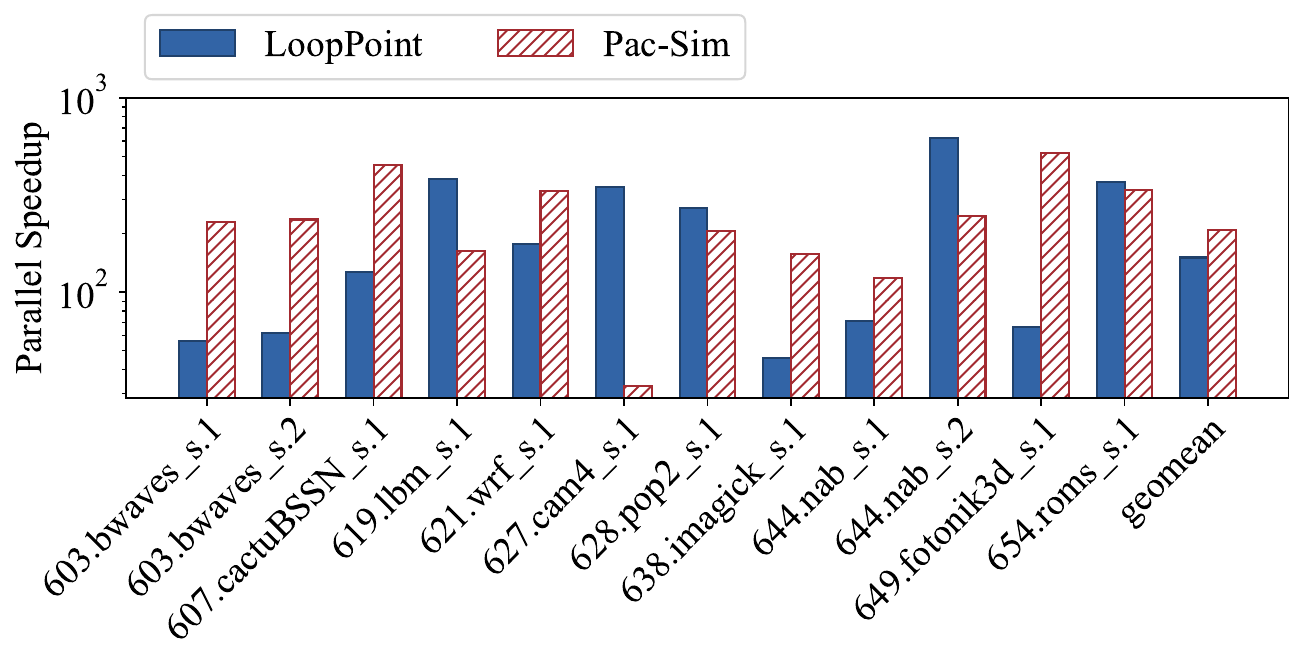}}
\hfill
    \subfloat[Serial Speedup\label{fig:sota:speedup}]{
    \includegraphics[width=0.48\linewidth]{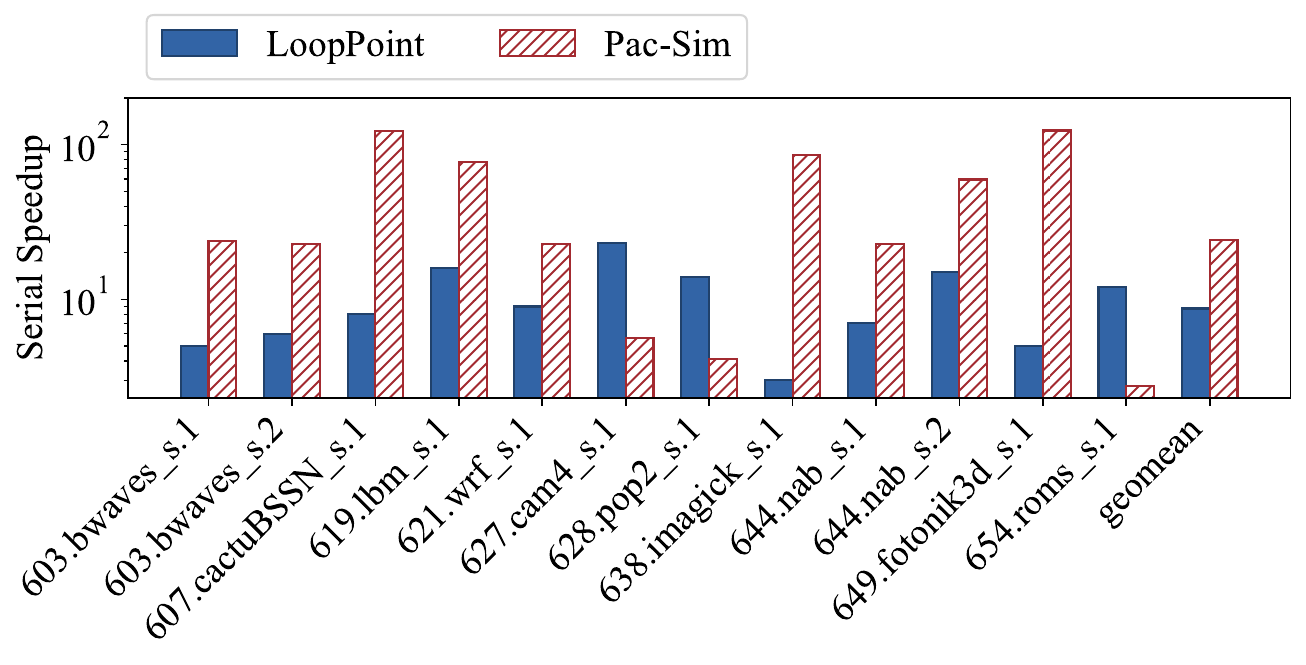}}
    \caption{The parallel and serial speedups of \projname{} are compared with that of LoopPoint for 8-threaded SPEC CPU2017 benchmarks using train inputs. For speedup calculations, the simulation walltime corresponding to \projname{} includes both online analysis and simulation time, whereas, for LoopPoint, we consider only the checkpoint simulation time, excluding the time required for offline profiling and checkpoint generation.}
    \label{fig:sota:both}
\end{figure*}

%% file: sections/7_results.tex
\begin{figure*}[!t]
    \centering
 \subfloat[Accuracy\label{fig:sota:arch:accuracy}]{    \includegraphics[width=.49\linewidth]{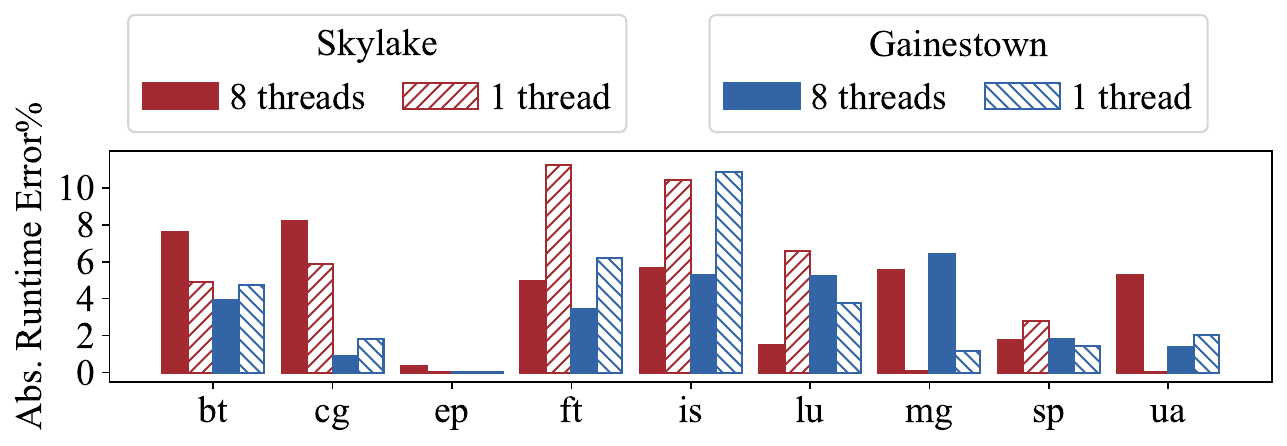}}    
\hfill
    \subfloat[Serial Speedup\label{fig:sota:arch:speedup}]{
    \includegraphics[width=.49\linewidth]{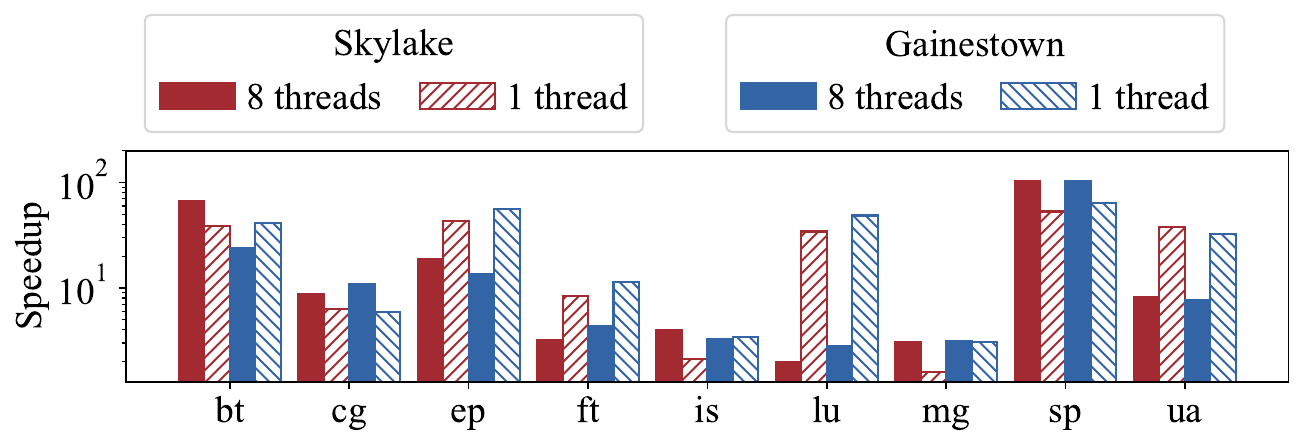}}
    \caption{The accuracy and serial speedup achieved for \projname{} methodology when simulated using Gainestown and Skylake architectures for NPB benchmarks with class A inputs running eight threads and one thread.}
     \label{fig:sota:arch:both}
\end{figure*}

In this section, we first present a comprehensive evaluation of \projname{}, comparing its efficacy with the current state-of-the-art. Additionally, we provide experimental evidence showing that \projname{} is indeed a hardware-independent methodology. Finally, we present case studies that demonstrate the applicability and effectiveness of \projname{} in estimating workload performance in dynamic, multi-threaded hardware and software environments.
Note that, throughout this paper, the term runtime refers to the simulated runtime of the application, whereas the term wall-time refers to the actual time taken by the simulator to finish the run.

\noindent\textbf{Evaluation metrics. } In order to evaluate the effectiveness of any simulation methodology, it is crucial to quantitatively measure its performance in terms of two critical metrics: \textit{accuracy} and \textit{speedup}. In our experiments, we define these metrics in the following manner:
    
\textit{Accuracy:} We assess the accuracy of our proposed methodology by comparing the simulation runtime obtained from the full simulation and the sampled simulation in terms of absolute runtime prediction error $\Delta_{time}$, which is defined as
    \begin{equation*}
        \Delta_{time} = \frac{|T_{full}-T_{sample}|}{T_{full}},
    \end{equation*}
    where $T_{full}$ represents the simulation runtime obtained from the full run, and $T_{sample}$ represents the simulation runtime extrapolated from the sampled simulation. 
   It is important to note that in our evaluation, we use the runtime (execution time as inferred from simulation) of the application as the performance metric to measure the accuracy of sampling. This is because time-per-program is the gold-standard performance measure, and IPC is not a valid performance metric for multi-threaded applications~\cite{2006alameldeen}.

\textit{Speedup:} 
In our experiments, we calculate the speedup by taking the ratio of the wall-clock time for the full simulation to that of the sampled simulation and the average speedup by computing the geometric mean of the speedups across all benchmarks. Serial speedup is defined as the speedup achieved when all representative regions are simulated sequentially, while parallel speedup is obtained when the representative regions are simulated in parallel, assuming infinite resources.

\subsection{Comparison with the State-of-the-Art}
\label{sec:accuracy:speedup}

In this section, we evaluate the performance of \projname{} in comparison to the state-of-the-art profile-driven sampled simulation methodology, LoopPoint~\cite{sabu2022looppoint}. While several other profile-driven methodologies exist, LoopPoint provides the benefit of being applicable across a variety of application and synchronization types. It has also been shown to outperform other multithreaded sampled simulation methodologies (such as BarrierPoint) in terms of speedup and accuracy, thus serving as a strong baseline for our evaluations.
We now report the results of our simulation experiments evaluating and comparing the performance of these two methodologies using the SPEC CPU2017 benchmarks.

\textbf{Accuracy.} Figure~\ref{fig:sota:err} shows a comparison of absolute runtime prediction errors for \projname{} and LoopPoint obtained for the 8-threaded SPEC CPU2017 benchmarks using train inputs. Our analysis reveals that, in most cases, \projname{} performs comparably with LoopPoint in predicting the runtime of the applications, with the individual errors differing by no more than 2 to 3\%.
The relatively higher errors for some applications, such as \texttt{619.lbm\_s.1} is because \projname{} relies on online extrapolation to estimate application performance using the limited profile data that is available from regions that have already been simulated. Whereas methodologies like LoopPoint rely on offline profiling and therefore utilize the information about the whole application.

\textbf{Speedup.}
\autoref{fig:sota:both} shows the speedup comparison of \projname{} and LoopPoint for the SPEC CPU2017 benchmarks using train inputs running eight threads.
Figure~\ref{fig:sota:parspeedup} shows the parallel speedup for which \projname{} outperforms LoopPoint in most cases (7 out of 12 benchmarks). The primary reason for this is that \projname{} uses smaller regions as compared to LoopPoint. Although Pac-Sim requires emulation of the entire application, the online analysis overhead is minimized, and therefore, the average parallel speedup for SPEC CPU2017 benchmarks (train inputs) using Pac-Sim is \avgspeedupparallel{}, which is larger than that obtained for LoopPoint (\avglooppointspeedupparallel{}). 

Figure~\ref{fig:sota:speedup} shows the serial speedup, and we observe \projname{} outperforms LoopPoint in most cases, attaining a maximum serial speedup of \MaxSpeedup{}. 
While the online analysis can introduce some runtime overheads, the performance advantages of \projname{} seem to outweigh these overheads in most cases. However, there are some cases where LoopPoint performs better than \projname{}, such as for \texttt{627.cam4\_s.1} and \texttt{628.pop2\_s.1} benchmarks in Figure~\ref{fig:sota:speedup}. This is mainly because \projname{} uses a small clustering threshold (0.05) for the online clustering in order to maintain high accuracy.

\begin{figure}[!t]
    \includegraphics[width=\linewidth]{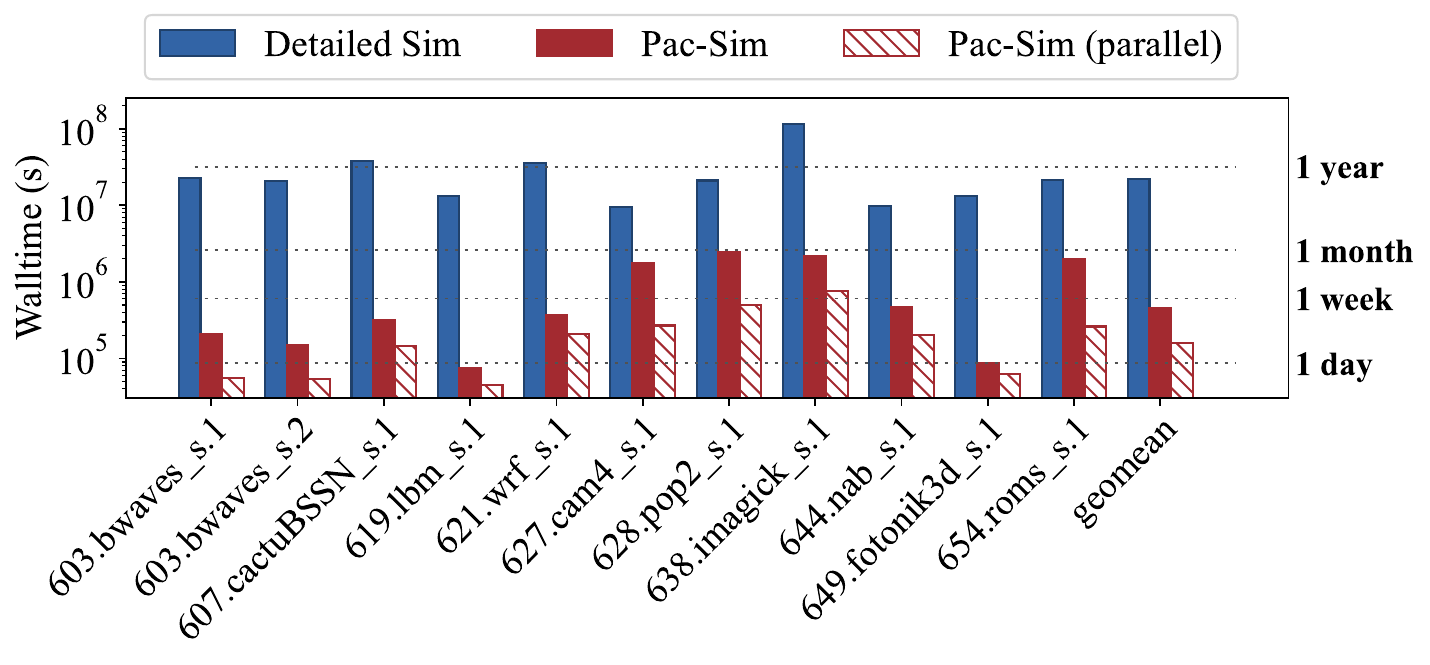}
    \caption{A comparison of the estimated walltime for fully detailed simulation and sampled simulation using the serial and parallel versions of \projname{} for 8-threaded SPEC CPU2017 benchmarks using ref inputs. The estimated walltime includes the time required for online analysis, warmup, and simulation.}
    \label{fig:ref:speedup}
\end{figure}
\textbf{Efficacy in Evaluating Realistic Workloads.} 
The full detailed simulation of SPEC CPU2017 benchmarks with reference inputs takes an extremely long time -- about a year on average using multi-core simulators like Sniper. Instead, we estimate their simulation walltime by considering the instruction count of the benchmark using reference inputs along with the average simulation rate of the benchmark using train inputs. The walltime of Pac-Sim includes the time required for online analysis and emulation of the entire workload along with the time for detailed simulation of the representative regions. \autoref{fig:ref:speedup} shows that Pac-Sim takes less than a week, on average, to run the entire application sequentially, while the parallel version of Pac-Sim takes about 1.8 days on average. In experiments where the microarchitecture structures like cache size are adjusted or when the application itself undergoes instruction-level modifications, it is necessary to regenerate the checkpoints. In such cases, Pac-Sim is more appropriate as LoopPoint takes 6.2 days on average (shown in \autoref{fig:resource}) to complete its preprocessing before simulation.

\textbf{Microarchitecture-agnostic sampling. }
In addition to achieving high accuracy and speedups, \projname{} also provides the advantage of being a microarchitecture-independent methodology. We experimentally demonstrate this by evaluating our methodology with two different processor configurations, namely the Gainestown and Skylake microarchitectures, for the NPB benchmarks that run using one thread and eight threads.
The accuracy and speedup numbers obtained in our experiments are plotted in Figure \ref{fig:sota:arch:accuracy} and Figure \ref{fig:sota:arch:speedup}, respectively.
From Figure~\ref{fig:sota:arch:accuracy}, we can observe that the absolute runtime errors estimated by \projname{} for all NPB applications are quite low (all under 8\%) and are similar for both these processor configurations (differing by 5\% at most).
Moreover, the speedups obtained for both configurations are similar for most benchmarks, as observed in Figure \ref{fig:sota:arch:speedup}. Hence, the choice of a target microarchitecture for evaluation does not affect the efficacy of \projname{}.

\begin{figure}[!t]
    \includegraphics[width=0.95\linewidth]{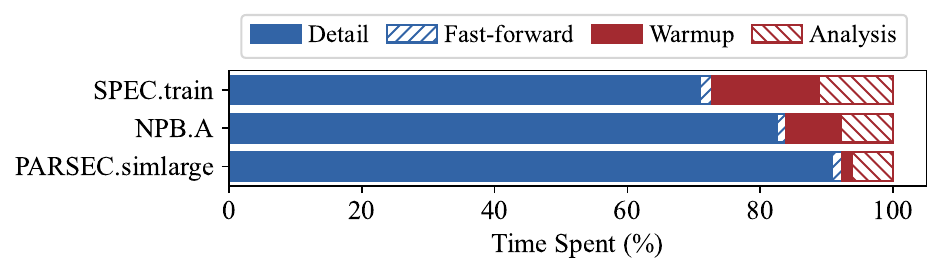}
    \caption{The graph shows the percentage of time that \projname{} spends at each phase during the sampled simulation of each benchmark suite (average across all benchmarks). The Analysis part includes online marker detection, region profiling, clustering, and prediction. 
    }
    \label{fig:timestack}
\end{figure}

\textbf{Wall-time Distribution.} We show the time spent by \projname{} in different stages of sampled simulation.
Figure~\ref{fig:timestack} shows the average time spent in the online analysis stage for NPB (class A inputs) and SPEC CPU2017 (train inputs) benchmarks is 7.88\% and 11.20\%, respectively. This is the result of the optimizations described in Section~\ref{sec:method}, which are applied to the analysis part. 
Moreover, \projname{} spends 8.25\% and 16.00\% of the execution time on warmup for NPB and SPEC CPU2017 benchmarks, respectively. This is because \projname{} needs to reconstruct the memory access patterns at the beginning of the detailed simulation of a region. Note that \projname{} brings down the time spent in profiling/analysis of the benchmarks significantly as compared to prior profile-driven methodologies for sampled simulation.

\subsection{Case Studies}
We showcase the versatility of \projname{} through several compelling case studies. Firstly, we demonstrate that our methodology remains agnostic to dynamic thread scheduling decisions made during runtime, highlighting its robustness and adaptability. Next, we provide examples of how \projname{} operates seamlessly in the presence of various runtime hardware events, further cementing its reliability. Finally, we exhibit the applicability of the proposed methodology in hardware-software co-design studies, showcasing its potential to facilitate more efficient and effective design processes.

\subsubsection{Dynamically Scheduled Software}

With the advent of multi-core and many-core processors, efficient parallel execution of dynamically scheduled multi-threaded applications has become crucial to the performance of modern computing systems. However, the non-determinism resulting from the execution of such applications on multi-core platforms often leads to notable performance variability across multiple runs. At the software level, such variability could arise from dynamic scheduling of system jobs, thread migrations, load balancing optimizations, or contention on shared resources at runtime.

\begin{table}[ht]
\caption{Table shows the IPC of \texttt{freqmine} benchmark from the PARSEC benchmark suite using \textit{simlarge} input for threads 0, \dots, 7. \projname{} shows the details of dynamically scheduled software whose IPC and thread mapping differ across two runs.}
    \centering
    \resizebox{\linewidth}{!}{%
    
    \begin{tabular}{c|cccccccc|c}
    \toprule
    Thread ID&0&1&2&3&4&5&6&7&Aggr.\\
    \midrule
    IPC$_{run1}$&0.15&0.09&1.75&\textbf{0.43}&\textbf{0.07}&0.07&0.10&0.09&2.75\\ 
    IPC$_{run2}$&0.15&0.09&1.76&\textbf{0.07}&\textbf{0.44}&0.07&0.09&0.10&2.76\\
    \bottomrule
     \end{tabular}
     }%
    \label{tab:dynamic}
\end{table}

Table~\ref{tab:dynamic} illustrates the thread-level differences in terms of IPC for two runs of the OpenMP-parallelized \texttt{freqmine} application from the PARSEC benchmark suite.  
There are variations in the per-thread IPCs between the two runs, particularly for thread IDs 3 and 4. To investigate the impact of these variations on conventionally used sampling techniques, we conducted two profiling runs of \texttt{freqmine} using LoopPoint. The experimental result revealed that 14\% of the regions were clustered differently for the second run as compared to the first. 
This presents a challenge for sampled simulation, which relies on profiling data from a prior execution to guide simulation in subsequent runs, as dynamic applications have variant profiling data across different executions.
To overcome these issues, \projname{} profiles and clusters the regions online and simulates them in the same run, thereby accounting for any performance variability that may occur at runtime.

\begin{figure}[t]
\centering
    \includegraphics[width=\linewidth]{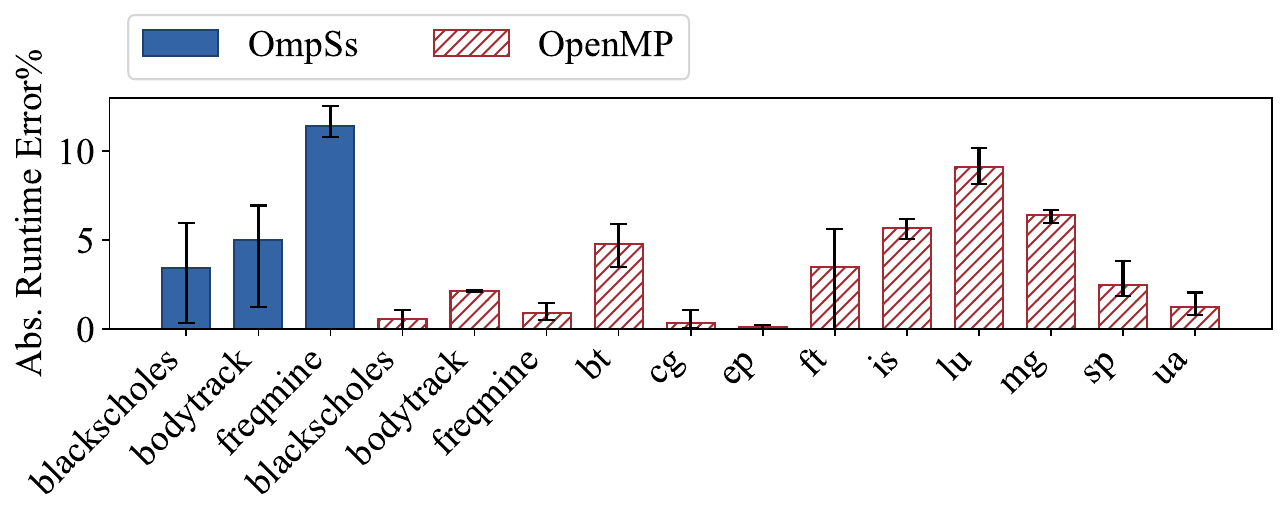}
    \caption{Figure shows the average error rates (from five different runs) and error bars in predicting the runtime of dynamically scheduled benchmarks. We use PARSEC benchmarks with the \textit{simlarge} input using OmpSs and OpenMP, and NPB benchmarks with \textit{class A} inputs using OpenMP runtime.
    }
  \label{fig:mtng:alldynamicsoftware}
\end{figure} 
      
To demonstrate the effectiveness of \projname{} in this regard, we now present an experimental study of dynamically scheduled multi-threaded versions of PARSEC with simlarge inputs and NPB with class A inputs.
While the per-thread behavior varies for dynamically scheduled applications, the global execution time and global IPC remain consistent across multiple runs. In Table~\ref{tab:dynamic}, we observe that while there are some variations in per-thread behavior, the aggregate IPCs across the two runs remain nearly unchanged.
Figure~\ref{fig:mtng:alldynamicsoftware} demonstrates the average runtime prediction errors of \projname{} simulating dynamically scheduled multi-threaded applications. We run the benchmarks multiple times in full detailed mode and using \projname{}. The errors are calculated by comparing the runtime obtained using \projname{} with the average runtime obtained from the full detailed simulations. 
The results show that \projname{} achieves a very low error in predicting the runtime of dynamically scheduled software (\DynamicAvgError{} on average). The benchmark, \texttt{freqmine}, which shows the
largest IPC variation maintains an average error of \DynamicFreqmError{}.
Moreover, \projname{} demonstrates speedups of up to \DynamicMaxSpeedup{} (\DynamicAvgSpeedup{} on average) for all dynamically scheduled benchmarks.

\subsubsection{Dynamic Hardware Events}

Dynamic event-based hardware optimizations help improve performance gains and energy efficiency in modern architectures. DVFS~\cite{eyerman2011fine,canturk2006analysis,kim2008system} is one of the most widely employed dynamic hardware event-based optimization techniques. It monitors core frequencies and load variations in order to match the system power consumption with the required level of performance by triggering voltage and frequency optimizations at runtime. These optimizations may lead to a diverse range of dynamic hardware states (i.e., core frequency, power configurations) over a given run, consequently resulting in a significant degree of performance variability for a given workload across different executions.

\projname{} deals with this performance variability by monitoring the simulated hardware events at runtime. While prior sampled simulation methodologies only support dynamic hardware events triggered at region boundaries to ensure that the hardware state remains constant for a given region, \projname{} supports hardware events at any time during the application execution. Each time an event occurs, \projname{} triggers a new region to ensure hardware state consistency within that region. The predictor then speculates the cluster ID of the next region and checks the execution history to determine whether similar regions (i.e., regions with the same cluster ID and hardware state) were previously encountered. If a similar region has been previously simulated in detail, the region is fast-forwarded; otherwise, a detailed simulation mode is triggered.

\begin{figure}[!t]
    \centering
    \subfloat[Aggregate GIPS from full detail simulation\label{fig:mtng:freq:full:ipc}]{
         \includegraphics[width=.95\linewidth]{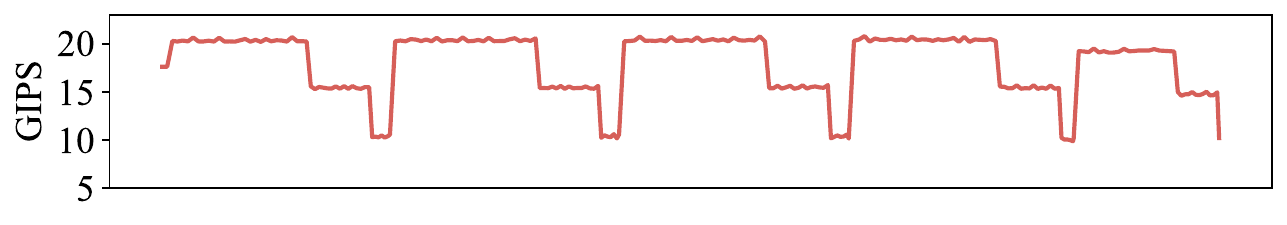}}
    \\
    \subfloat[Reconstructed GIPS using \projname{}\label{fig:mtng:freq:ipc}]{ 
         \includegraphics[width=.95\linewidth]{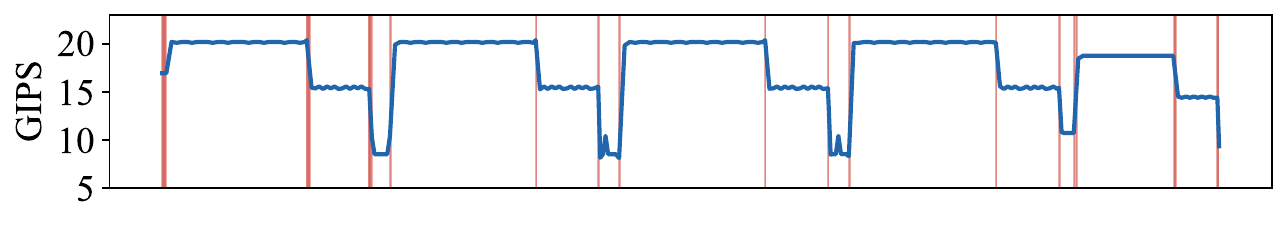}}
    \\
    \subfloat[CPU Frequency\label{fig:freqs}]{
       \includegraphics[width=.95\linewidth]{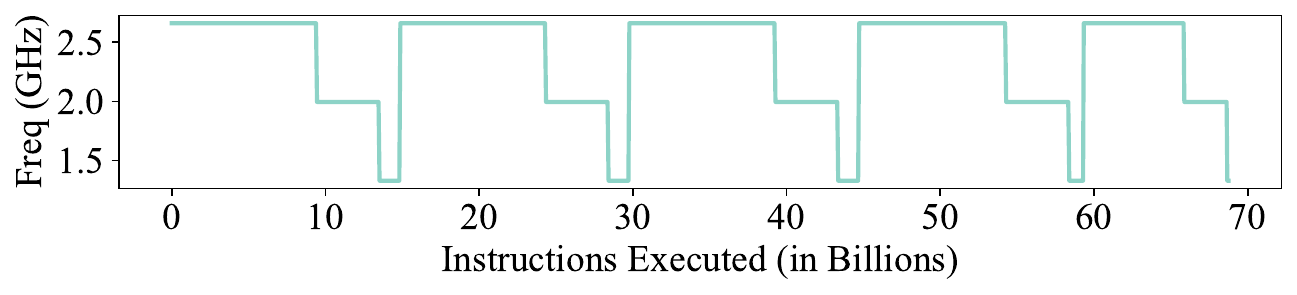}}
    \caption{The aggregate giga (billion) instructions per second (GIPS)
    of the full run (a), reconstructed GIPS using \projname{} (b), and the varying CPU frequency for all CPUs (c) \texttt{644.nab\_s.1} benchmark with train inputs running 8 threads. The shaded regions in (b) represent the regions simulated in detail. The figures share the same x-axis.}
 
    \label{fig:case-study-gips}
\end{figure}

We now present an experimental study demonstrating the effectiveness of \projname{} in handling the variability caused by dynamic hardware events by specifically considering the case of DVFS-optimized workloads. In our experiments, we evaluate the performance of the benchmarks by comparing the results of \projname{} with the baseline while changing the frequency at predetermined intervals; however, just like in actual DVFS-optimized executions, the information on the frequency changes is not available to the simulator a priori. In order to evaluate the performance of \projname{}, we consider a DVFS scenario in which the processor frequency $\mathcal{f}$ switches among a fixed range of values, i.e., $\mathcal{f} \in \{$1.33 GHz, 2.00 GHz, 2.66 GHz$\}$ as shown in Figure~\ref{fig:freqs}.

For this scenario, we measure the aggregate giga/billion instructions per second (GIPS) values obtained from both the full detailed simulation and \projname{} over the entire execution. The findings of our experiment are presented in Figure~\ref{fig:case-study-gips}. We observe that the GIPS values obtained from both the full simulation (Figure~\ref{fig:mtng:freq:full:ipc}) and \projname{} (Figure~\ref{fig:mtng:freq:ipc}) exhibit a great deal of similarity, indicating \projname{}'s effectiveness in estimating the performance of a dynamically optimized workload with a high level of accuracy. 
Furthermore, our findings reveal that \projname{} simulates only a small fraction of the entire application in detail (depicted by shaded regions in Figure~\ref{fig:mtng:freq:ipc}). Notably, most of the detailed simulation occurs either at points of change in the phase behavior of the application or hardware states.
This demonstrates that \projname{} can use this information to identify a minimal representative subset for applications using online analysis.

\subsubsection {Hardware-Software Co-design}

\begin{figure}[!t]
    \centering
    \includegraphics[width=0.99\linewidth]{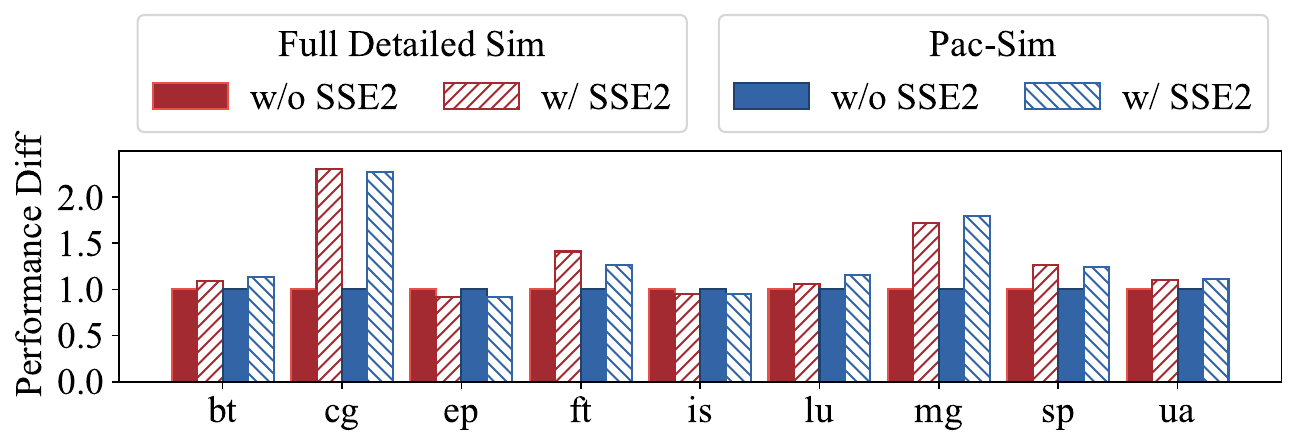}
    \caption{
    The figure shows the absolute difference in performance (in terms of runtime) for NPB benchmarks using class A inputs and 8 threads with (w/) and without (w/o) SSE2 simulated in detailed mode and with \projname{}.
    }
    \label{fig:hardware-software-codesign}
\end{figure}

Hardware-software co-design is an emerging field of study that optimizes the system performance by concurrently designing the compiler and hardware components of a system to exploit the synergy between the two. Prior works \cite{noreba2021, zeng2021turnpike, eyerman2021enabling} have investigated several directions in this context. To identify the most effective strategies, hardware-software co-design research relies on fast and accurate architectural simulation methodologies to explore the design space efficiently. However, among existing methodologies, the profile-driven methodologies \cite{2002Sherwood, 2014carlson} incur significant profiling and preprocessing costs, as shown in Figure~\ref{fig:resource}, whereas the statistical sampling methodologies \cite{2003Wunderlich, PCantorSim} (which don't rely on preprocessing) have low speedups.

\projname{} addresses these issues by sampling and analyzing the regions online. Thus, it incurs no additional profiling cost if new compilers are used or new applications are generated, enabling fast and efficient exploration of hardware-software co-design space. 
To demonstrate the effectiveness of \projname{} in this regard, we now present a performance evaluation study of the NPB benchmarks under different compiler optimization techniques.
We study the impact of SIMD (Single Instruction, Multiple Data) optimizations on the generated binaries using both \projname{} and full detailed simulations.
SIMD-enabled processors are equipped with special-purpose registers that can simultaneously load multiple machine words and perform operations on them in parallel in order to improve processor performance. For instance, the Streaming SIMD Extensions 2 (SSE2) instruction set uses 128-bit XMM registers to process packed data elements at once.

In our experiments, we measure the performance improvement (in terms of runtime) obtained by enabling SSE2 and compare it against the baseline. 
The results of our simulations are graphed in Figure~\ref{fig:hardware-software-codesign}.
We observe that the average difference in the performance improvements obtained from full detailed mode and \projname{} is \AVGSPEEDUPDIFF{}. 
Specifically, \projname{} reveals the performance effects of SIMD instructions.  
For example, some benchmarks achieve a significant speedup over the baseline as these applications meet the \texttt{icc} vectorization criteria~\cite{deilmann2012guide}. \texttt{ft} calculates a 3D fast Fourier transform, and its innermost loop consists of multiply-add statements with contiguous memory accesses and no data dependency.
On the other hand, \texttt{is}, which uses the quick sort algorithm, is hard to vectorize. 
The SIMD overheads resulting from register transfer costs exacerbate the overall application performance.

%% file: sections/8_related_work.tex
We have discussed the most relevant previous works in Section~\ref{sec:motivation}. Sampled simulation has been an active research area for several decades, and several techniques were proposed~\cite{2002Sherwood,2003Wunderlich,hassani2016livesim,sandberg2015fsa,2006wenisch,argollo2009cotson,2013carlson,2013ardestani,2014carlson,2016grass,sabu2022looppoint} in this direction for different workload classes primarily for the reduction of simulation time and resources. 
In order to simulate the identified sample correctly, it is important to reconstruct the microarchitectural state of the system using techniques like statistical warming~\cite{haskins2003memory,barr2005accelerating,eeckhout2005blrl,nikoleris2019micro}, checkpoint-based warming~\cite{nikoleris2016coolsim,van2006efficient}, or by enabling functional simulation.
Analytical modeling is yet another solution to evaluate a complex workload quickly. Prior works proposed analytical models to derive the performance of processors~\cite{eyerman2009mechanistic,depestel2019rrppomwomp}, cache miss rates~\cite{eklov2010statstack}, branch miss rates~\cite{de2015micro}, DVFS performance~\cite{akram2016dvfs}, etc. However, analytical performance modeling 
can be limited in supporting new designs, requiring new models for each.  

%% file: sections/90_conclusion.tex
This work proposes a novel sampled simulation methodology and infrastructure called \projname{}. 
The work focuses on what is needed to simulate dynamic software that responds to workload- and run-time-specific execution conditions.
\projname{} is the first, to the best of our knowledge, to propose a sampling solution that simulates these dynamic conditions in both a fast (up to \MaxSpeedupNaturalParallel{} speedup, \AvgSpeedupNaturalParallel{} on average) and accurate way (average errors of
\AVGErrorNatural{} and \DynamicAvgError{} for statically and dynamically scheduled benchmarks, respectively).